\documentclass[a4paper]{modifiedjpconf}
\usepackage{amsmath, amsfonts, amsthm, amssymb, graphicx, slashed}

\newcommand{\nc}{\newcommand}
\nc{\rnc}{\renewcommand}
\nc{\pre}{{\it Preprint }}
\rnc{\d}{\mathrm{d}}
\nc{\D}{\partial}
\rnc{\t}{\tau}
\nc{\K}{\kappa}
\nc{\g}{\gamma}
\nc{\lrarrow}{\leftrightarrow}
\nc{\rg}{\sqrt{g}}
\rnc{\[}{\begin{equation}}
\rnc{\]}{\end{equation}}
\nc{\bea}{\begin{eqnarray}}
\nc{\eea}{\end{eqnarray}}
\nc{\nn}{\nonumber}
\rnc{\(}{\left(}
\rnc{\)}{\right)}
\nc{\ep}{\epsilon}
\nc{\tto}{\rightarrow}
\rnc{\inf}{\infty}
\rnc{\Re}{\mathrm{Re}}
\rnc{\Im}{\mathrm{Im}}
\nc{\ie}{{\it i.e.~}}
\nc{\iec}{{\it i.e.,~}}
\nc{\vphi}{\varphi}      
\nc{\tq}{\bar{q}}
\nc{\tK}{\bar{\K}}
\nc{\tV}{\bar{V}}
\nc{\z}{\zeta}
\nc{\Z}{\mathcal{Z}}
\nc{\<}{\langle}
\rnc{\>}{\rangle}
\rnc{\O}{\mathcal{O}}

\begin{document}

\title{The Holographic Universe}

\author{P L McFadden and K Skenderis}

\address{Institute for Theoretical Physics, Valckenierstraat 65, 1018XE Amsterdam, the Netherlands.}
\ead{P.L.McFadden@uva.nl, K.Skenderis@uva.nl}


\begin{abstract}

We present a holographic description of four-dimensional
single-scalar inflationary universes in terms of a three-dimensional
quantum field theory (QFT). The holographic description correctly reproduces
standard inflationary predictions in their regime of applicability.
In the opposite case, wherein gravity is strongly
coupled at early times, we propose a holographic description in terms
of perturbative QFT and present models capable of satisfying the current
observational constraints while exhibiting a phenomenology
distinct from standard inflation. This provides
a qualitatively new method for generating
a nearly scale-invariant spectrum of primordial cosmological perturbations.

\end{abstract}

\section{Introduction.}

The notion of holography emerged from black hole physics as an answer to
the question: why is the entropy of a black hole proportional to the
area of its horizon?
Since entropy is an extensive quantity, one would have expected
the entropy to be instead proportional to the volume the black hole occupies.
Typically, entropy is a measure of the number of degrees
of freedom.  The scaling of the gravitational entropy
has thus been taken as an indication of a new fundamental principle,
the holographic principle, that underlies any quantum theory of gravity.
More precisely, holography states that
any quantum gravitational system in $(d+1)$ dimensions should
have a dual description in terms of a QFT without gravity in one dimension less \cite{'tHooft:1993gx}.
Indeed, if gravity is holographic this would
explain the scaling behavior of the black hole entropy, since
the entropy of a QFT scales like the volume,
which is the same as the area in one dimension higher.

Holography provides a new paradigm for physical reality,
the consequences of which we are only just beginning to
comprehend. According to this picture, one of the
macroscopic dimensions of spacetime and one of the
forces in the universe, namely gravity,
are emergent phenomena in an underlying lower-dimensional QFT.
Concrete realizations of holography have been found
in string theory, and a precise holographic
dictionary was established shortly thereafter
\cite{Maldacena:1997re,Gubser:1998bc,Witten:1998qj}.
To date, almost all such realizations involve spacetimes with
a negative cosmological constant. The arguments
that led to the holographic principle apply more generally, however,
and suggest that one should be able to establish a
holographic dictionary that applies to our
own universe.  The purpose of the work presented
in \cite{McFadden:2009fq}, and further discussed here,
is to propose a concrete holographic framework
that applies to our own universe, and
in particular to its cosmological evolution.
More precisely, we will describe how to set up holography for
inflationary cosmology.

Any holographic proposal for cosmology should specify
(i) what the dual QFT is, and (ii) how it can be used to
compute cosmological observables. Having defined such
a duality, the new description should recover established results
in their regime of applicability. Indeed, we will see that our holographic
models correctly reproduce standard inflationary results
when standard inflation is applicable, namely,
under the assumption that gravity was weakly coupled
at early times.  Yet perhaps more importantly, our holographic approach
also gives a qualitatively new method for generating
a nearly scale-invariant spectrum when gravity was
{\it strongly} coupled at early times. As we will discuss in the following,
there exist holographic models that are capable of satisfying all current
observational constraints while exhibiting a phenomenology
distinct from standard inflation.

Over the last two decades striking new observations have transformed
cosmology from a qualitative to quantitative science \cite{Komatsu:2008hk}.
These observations reveal a spatially flat
universe, endowed with small-amplitude primordial perturbations that
are approximately Gaussian and adiabatic with a nearly scale-invariant
spectrum.  This data is consistent with the generic predictions of
inflationary cosmology and set inflation as the leading theoretical
paradigm for the initial conditions of Big Bang cosmology. Yet
inflation, despite these successes, is still unsatisfactory in
a number of ways: it generically requires fine tuning and there are
trans-Planckian issues and questions about the initial conditions
for inflation \cite{Turok:2002yq}.

With future observations promising an unprecedented era
of precision cosmology, it becomes imperative that inflation
is embedded in a UV complete theory (indeed there is
increasing amount of effort devoted to embedding inflation in
string theory), and it is
also important that alternative scenarios are developed.
The holographic approach that we undertake
provides both. Firstly, the holographic models we will discuss
here are three-dimensional super-renormalizable theories
and thus are UV complete. Secondly, holographic
dualities are strong/weak coupling dualities meaning that in the
regime where one description is weakly coupled, the other is
strongly coupled. This provides an arena for constructing new models
with intrinsic strong-coupling gravitational dynamics at early times
that possess only a weakly coupled three-dimensional QFT description, and
are thus outside the class of model described by standard inflation.
As we will see, such models may lead to qualitatively different predictions for
the cosmological observables that will be measured in the near future.
Furthermore, quite apart from the fresh perspective on early universe cosmology
such an approach offers, there are also a number of more pragmatic reasons for
developing a holographic framework for cosmology: uncovering
the structure of three-dimensional QFT in cosmological observables
brings in new intuition about their structure and may lead to more
efficient computational techniques, {\it cf.}~the computation of
non-Gaussianities in \cite{Maldacena:2002vr}.

The holographic
description we propose uses the one-to-one correspondence between
cosmologies and domain-wall spacetimes discussed in
\cite{Cvetic:1994ya, Skenderis:2006jq} and assumes that the standard
gauge/gravity duality is valid. More precisely, the steps involved are
illustrated in Fig.~1.  The first step is to map any given
inflationary model to a domain-wall spacetime.  For inflationary
cosmologies that at late times approach either a de Sitter spacetime or a power-law
scaling solution\footnote{This era should then be followed
by a hot big bang cosmology, as in standard discussions. Here we
only discuss the very early universe, \iec the times when the primordial
cosmological perturbations were generated (the inflationary epoch).},
the corresponding domain-wall solutions describe
holographic renormalization group flows.
For these cases there is an
operational gauge/gravity duality, namely one has a dual description
in terms of a three-dimensional QFT.  Now, the map between cosmologies
and domain-walls may equivalently be expressed entirely in terms of
QFT variables, and amounts to a certain analytic continuation of
parameters and momenta.  Applying this analytic continuation, we obtain
the QFT dual of the original cosmological spacetime.

We shall call the resulting theory a `pseudo'-QFT because we currently
have only an operational definition of this theory. Namely, we do the
computations in the QFT dual to the corresponding domain-wall
and then apply the analytic continuation.  Perhaps a more fundamental
perspective is to consider the QFT action, with complex parameters and
complex fields as the fundamental objects, and then to consider the
results on different real domains as applicable to either domain-walls
or cosmologies.  Note that the supergravity embedding of the
domain-wall/cosmology correspondence discussed in
\cite{Bergshoeff:2007cg} works in precisely this way.

This paper is organized as follows. In the next section we
discuss the domain-wall/cosmology correspondence. Then, in
Section \ref{sec:cosmo_pert}, we discuss the cosmological
observables that we would like to compute holographically,
and in Section \ref{sec:Hol_an} we present the holographic
analysis. In Section \ref{sec:anal_cont}, we discuss the
analytic continuation to the pseudo-QFT and in
 Sections \ref{sec:beyond} and \ref{sec:holo_pheno}
we discuss the new models that are strongly coupled at early times
but have a weakly coupled QFT description.

\begin{figure}[tr]
\includegraphics[width=0.65\textwidth]{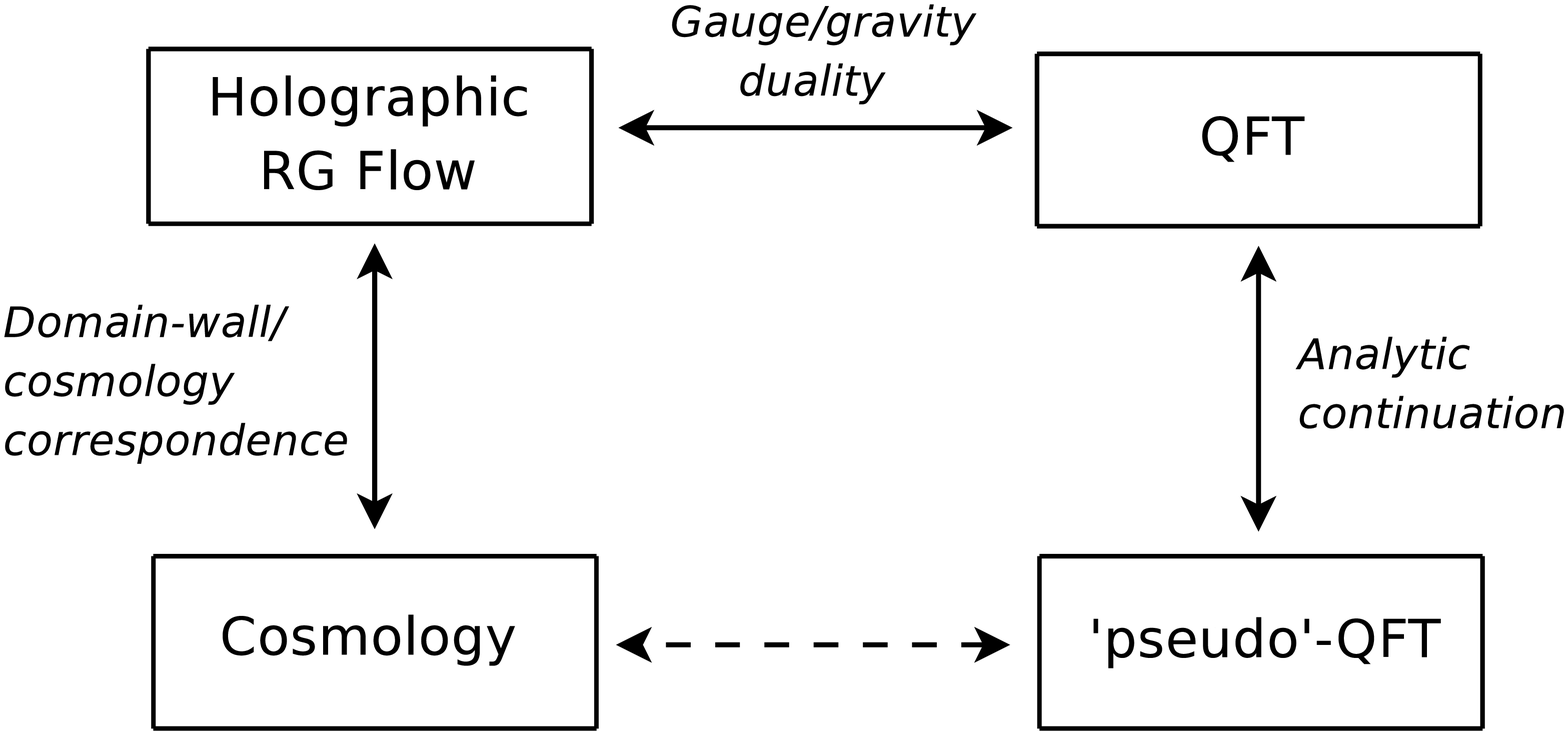}
\hspace{2pc}
\begin{minipage}[b]{11pc}
\caption{The `pseudo'-QFT dual to inflationary cosmology is operationally defined
using the correspondence of cosmologies to domain-walls and standard gauge/gravity duality.}
\end{minipage}
\end{figure}

\section{Domain-wall/cosmology correspondence.}

We explain in this section the lefthand vertical line in Fig.~1,
namely, the correspondence between
cosmologies and domain-wall spacetimes.
For simplicity, we focus on spatially flat universes equipped with a
single minimally coupled scalar field, but the results can be extended
to more general cases ({\it eg.}, non-flat, multi-scalar,
non-canonical kinetic terms, {\it etc}).
The metric and scalar field for the unperturbed solution are given by
\[
\d s^2 = \eta \d z^2+a^2(z) \d \vec{x}^2, \qquad
\Phi = \vphi(z), \label{back}
\]
where $\eta = -1$ in the case of cosmology, in which case $z$ is the time coordinate,
and $\eta = +1$ in the case of domain-wall solutions\footnote{
The name `domain-wall spacetime' dates back to earlier work featuring
solutions of this form that interpolate between two stationary points of the scalar field 
potential, one at $z=+\infty$ and another at $z=-\infty$.  In the present context the name 
is somewhat misleading, however, since we consider
only the $z>0$ part of the geometry. We will nevertheless stick with this terminology as it is 
standard usage in high-energy physics.} in which case $z$ is the radial coordinate.
We take the domain-wall to be Euclidean for later convenience.
A Lorentzian domain-wall may be obtained by continuing
one of the $x^i$ coordinates to become the time coordinate \cite{Skenderis:2006jq}.
The continuation to a Euclidean domain-wall is convenient, however, because the QFT vacuum state 
implicit in the Euclidean formulation maps to the Bunch-Davies vacuum on the cosmology side.
Other choices of cosmological vacuum require considering the boundary QFT
in different states, as may be accomplished using the real-time formalism of  \cite{Skenderis:2008dh}.

With the appropriate choice of $a(z)$ and $\vphi(z)$,
the configuration in (\ref{back}) solves the field equations that follow
from the action 
\[
\label{Action}
S = \frac{\eta}{2\K^2}\int \d^4 x\sqrt{|g|}\, [-R+(\D\Phi)^2+2\K^2 V(\Phi)],
\]
where $\K^2 = 8\pi G$ and we are taking the scalar field $\Phi$ to be dimensionless.
Note that when $\eta=-1$, the kinetic terms
have the appropriate signs for a mostly plus Lorentzian signature metric
and when $\eta=+1$, the kinetic terms have the correct sign for Euclidean
signature. Had we expressed both actions in the same signature metric,
they would differ only in the sign of the potential. It follows that
for every flat FRW solution of a model with potential $V$ there is a corresponding
domain-wall solution of a model with potential $-V$
\cite{Cvetic:1994ya,Skenderis:2006jq}.

For background solutions in which the evolution of the scalar field is
(piece-wise) {\it monotonic}, $\vphi(z)$ can be inverted to give
$z(\vphi)$ permitting the Hubble rate $H=\dot{a}/a$ to be re-expressed as
$
H(z) = - (1/2) W(\vphi),
$
where $W(\vphi)$ is known as the `fake superpotential'.
In this case, the complete equations of motion for the background
take the simple form
\[
\label{firstorder2}
\frac{\dot{a}}{a} = -\frac{1}{2}W, \qquad
\dot{\vphi} = W_{,\vphi}, \qquad 2\eta \K^2 V = (W_{,\vphi})^2-\frac{3}{2} W^2.
\]
This first-order formalism goes back to the work of \cite{Salopek:1990jq} (for cosmology),
where it was obtained by application of the Hamilton-Jacobi method.
In \cite{Skenderis:2006jq} this formalism was linked to the notion of (fake) (pseudo-) supersymmetry.

We will now extend the correspondence to encompass linear perturbations
around the background solution.  The linearly perturbed metric takes the general form
\begin{align}
\d s^2 &= \eta [1+2\phi(z,\vec{x})]\d z^2+2a^2(z)[\partial_i \nu(z,\vec{x})+\nu_i(z,\vec{x})]\d z\d x^i
+a^2(z)[\delta_{ij}+h_{ij}(z,\vec{x})]\d x^i\d x^j, \nn \\
\Phi &= \vphi(z)+\delta\vphi(z, \vec{x}), \label{lin_fl}
\end{align}
where $\nu_i$ is transverse. 
The spatial metric perturbation $h_{ij}$ may be decomposed as
\[
\label{metric_decomp}
h_{ij}(z, \vec{x}) = -2\psi(z, \vec{x})\delta_{ij}+2\partial_i\partial_j\chi(z, \vec{x})+2\partial_{(i}w_{j)}(z,\vec{x})+
\g_{ij}(z, \vec{x}),
\]
where $\omega_i$ is transverse and $\g_{ij}(z, x)$ is transverse traceless.
The metric perturbations may then be combined into the gauge-invariant combinations
\begin{align}
\label{zeta}
\zeta &= \psi+(H/\dot{\vphi})\delta\vphi, \\
\hat{\phi} &= \phi-(\delta\vphi/\dot{\vphi})\dot{},\\
\hat{\nu} &= \nu-\dot{\chi}-\eta(\delta\vphi/a^2\dot{\vphi}),\\
\hat{\nu}_i &= \nu_i-\dot{\omega}_i.
\end{align}
Physically, $\zeta$ represents the curvature perturbation on comoving
hypersurfaces and has the useful property that it tends to a constant on superhorizon scales\footnote{
Following the end of inflation, the superhorizon value of $\zeta$ then remains constant until horizon re-entry, irrespective of
the dynamics of the intervening evolution, provided that no entropy perturbations are produced \cite{Bardeen:1983qw, Mukhanov:1990me}.}.

The equations of motion for cosmological perturbations have been worked out long ago, see \cite{Mukhanov:1990me} and references therein,
while the corresponding analysis for domain-walls may be found in
\cite{DeWolfe:2000xi,Bianchi:2001de, Papadimitriou:2004rz}.
In the present case, there are two independent perturbations represented by
$\zeta$ and $\g_{ij}$, since the Hamiltonian and momentum constraints are equivalent to
\[
\label{constraints}
\hat{\phi}=-\frac{\dot{\zeta}}{H}, \qquad \hat{\nu}=-\frac{\eta\zeta}{a^2H}+\frac{\ep\dot{\zeta}}{q^2}, \qquad \hat{\nu}_i=0,
\]
where $\vec{q}$ is the comoving wavevector of the perturbations,
and the background quantity $\ep(z)$ is defined as
$\ep = -\dot{H}/H^2 =2(W_{,\vphi}/W)^2$.
(In standard inflation $\ep$ would be the usual slow-roll parameter; however,
we do not assume slow roll here).
From the remaining Einstein equations, one then finds the following equations of motion for $\zeta$ and $\g_{ij}$:
\begin{align}
\label{pert_eom}
0&= \ddot{\zeta}+(3H+\dot{\ep}/\ep)\dot{\zeta}- \eta a^{-2} q^2\zeta, \nn \\
0&= \ddot{\gamma}_{ij}+3H\dot{\g}_{ij}-\eta a^{-2}q^2\g_{ij}.
\end{align}

Defining now the analytically continued variables $\tK$ and $\tq$ according to
\[
\label{a/c}
\tK^2  = -\K^2, \qquad \tq=-iq\, ,
\]
it is easy to see that a perturbed cosmological solution written in terms of the variables $\K$ and $q$ continues to a perturbed Euclidean domain-wall solution expressed in terms of the variables $\tK$ and $\tq$.  Note that (\ref{pert_eom}) only requires $\tq^2 = - q^2$ and in (\ref{a/c}) we made a choice
of a branch cut in the function $q = \sqrt{q^2}$ (for reasons to be explained in the next section).
As it is clear from (\ref{firstorder2}), continuing $\K$ is equivalent to continuing $V$.  Here we prefer to continue $\K$  since, as we will see, the former has a clear interpretation in terms of dual QFT variables.

We have thus established that the correspondence between cosmologies
and domain-walls holds, not only for the background solutions, but
also for linear perturbations around them.  This is the basis for the
relation between power spectra and holographic 2-point functions, to
be discussed momentarily.  The argument can be generalized to
arbitrary order to relate non-Gaussianities to holographic
higher-point functions \cite{to_appear}.

\section{Cosmological  observables.} \label{sec:cosmo_pert}

In the inflationary paradigm, cosmological perturbations
originate on sub-horizon scales as quantum fluctuations of the vacuum.
Quantizing the perturbations in the usual manner, one finds the scalar and tensor
superhorizon power spectra
\begin{align}
\Delta^2_S(q)& = \frac{q^3}{2\pi^2}\< \z(q)\z(-q)\>= \frac{q^3}{2\pi^2}|\z_{q(0)}|^2, \nn \\
\Delta^2_T(q) &= \frac{q^3}{2\pi^2}\< \g_{ij}(q)\g_{ij}(-q)\>=\frac{2 q^3}{\pi^2}|\g_{q(0)}|^2,
\end{align}
where $\g_{q(0)}$ and $\z_{q(0)}$ are the constant late-time values of the cosmological mode functions $\g_q(z)$ and $\z_q(z)$.

The mode functions are themselves solutions of the classical equations of motion (\ref{pert_eom})
(setting $\g_{ij}=\g_q e_{ij}$, for some time-independent polarization tensor $e_{ij}$).
To select a unique solution for each mode function we
impose the Bunch-Davies vacuum condition $\zeta_q, \g_q \sim \exp(-iq\t)$
as $\t \tto - \inf$, where the conformal time $\t = \int^z \d z'/a(z')$.
The normalization of each solution (up to an overall phase) may then be fixed by imposing the canonical commutation relations for the corresponding quantum fields.
This leads to the Wronskian conditions,
\[
\label{Wronsk}
i = \z_q \Pi^{(\z)*}_q - \Pi^{(\z)}_q \z_q^*, \qquad
i/2 = \g_q \Pi^{(\g)*}_q - \Pi^{(\g)}_q \g_q^*,
\]
where $\Pi^{(\zeta)}_q= 2 \ep a^3\K^{-2} \dot{\z_q}$ and $\Pi^{(\g)}_q = (1/4)a^3\K^{-2}\dot{\g}_q$ are the canonical momenta associated with each mode function, and we have set $\hbar$ to unity.

To make connection with the holographic analysis to follow,
we introduce the linear response functions $E$ and $\Omega$ satisfying
\[
\label{response_fns}
\Pi^{(\zeta)}_q=\Omega \,\zeta_q, \qquad  \Pi^{(\g)}_q= E \,\g_q.
\]
These quantities are well-defined since we have already selected a unique solution for each mode function.
Substituting these definitions into the Wronskian conditions,
which are valid at all times, the cosmological power
spectra may be re-expressed as
\[
\label{cosmo_result}
\Delta^2_S(q) = \frac{-q^3}{4\pi^2 \Im \Omega_{(0)}(q)}, \quad
\Delta^2_T(q) = \frac{-q^3}{2\pi^2 \Im E_{(0)}(q)},
\]
where $\Im \Omega_{(0)}$ and $\Im E_{(0)}$ are the constant late-time values of the
imaginary part of the response functions (more precisely, the subscript indicates
that this is the part of the response
function that is invariant under dilatations, see the discussion at the end of subsection \ref{sbsec:basics}).
We will see shortly how the response functions also give the 2-point function of the pseudo-QFT.

Let us now consider the corresponding domain-wall solution obtained by the applying the continuation (\ref{a/c}).
The early-time behavior $\sim \exp(-iq\t)$ of the cosmological perturbations maps to the exponentially decaying behavior $\sim \exp(\tq \t)$ in the interior of the domain-wall ($\t \tto -\inf$).
Such regularity in the interior is a prerequisite for holography, explaining our choice of sign in the continuation of $q$.

The domain-wall response functions $\bar{E}$ and $\bar{\Omega}$
\cite{Papadimitriou:2004rz} are defined analogously to (\ref{response_fns}),
namely
\[
\label{dw_response_fns}
\bar{\Pi}^{(\z)}_{\tq} =-
\bar{\Omega}\, \z_{\tq} \qquad
\bar{\Pi}^{(\g)}_{\tq} =-
\bar{E}\, \g_{\tq},
\]
where $\bar{\Pi}^{(\z)}_{\tq} = 2\ep a^3\tK^{-2} \dot{\zeta}_{\tq}$ and
$\bar{\Pi}^{(\g)}_{\tq} = (1/4)a^3\tK^{-2}\dot{\g}_{\tq}$
are the radial canonical momenta. The minus sign in (\ref{dw_response_fns}) are inserted so that
\[
\bar{\Omega}(-iq) =\Omega(q), \qquad \bar{E}(-iq) =E(q).
\]
By choosing the arbitrary overall phase of the cosmological perturbations appropriately, we may ensure that the domain-wall perturbations are everywhere real.  The domain-wall response functions are then purely real, while their cosmological counterparts are complex.

\section{Holographic analysis.} \label{sec:Hol_an}

In this section we briefly review relevant material from
gauge/gravity duality, corresponding to the upper horizontal
line in Fig.~1. There are two classes of domain-wall solutions for
which holography is well understood:

\paragraph{Asymptotically AdS domain-walls.}

In this case the solution behaves asymptotically as
\[
a(z) \sim e^{z}, \qquad \vphi \sim 0 \qquad {\rm as} \quad z \to \infty.
\]
The boundary theory has a UV fixed point which corresponds to the bulk AdS critical point.
Depending on the rate at which $\vphi$ approaches zero as $ z \to \infty$,
the QFT is either a deformation of the conformal field theory (CFT), or else the CFT in a state in which the dual scalar operator acquires a nonvanishing vacuum expectation value (see \cite{Skenderis:2002wp} for details).
Under the domain-wall/cosmology correspondence, these solutions are mapped to
cosmologies that are asymptotically de Sitter at late times.

\paragraph{Asymptotically power-law solutions.}

In this case the solution behaves asymptotically as
\[
\label{power_law}
a(z) \sim (z/z_0)^n, \quad \vphi \sim \sqrt{2 n} \log (z/z_0) \quad {\rm as} \quad z \to \infty,
\]
where $z_0=n{-}1$.
This case has only very recently been understood \cite{Kanitscheider:2008kd}.
 For $n=7$ the asymptotic geometry is the near-horizon limit of a stack of D2 brane solutions.
In general, these solutions describe QFTs with a dimensionful coupling
constant in the regime where the dimensionality of the coupling constant drives the dynamics.
Under the domain-wall/cosmology correspondence, these solutions are mapped to cosmologies that
are asymptotically power-law at late times.

\subsection{Basics of holography.} \label{sbsec:basics}

Gauge/gravity duality is an exact equivalence between a bulk gravitational
theory and a boundary QFT. Typically, the boundary QFT is a gauge theory that admits a large $N$ expansion.
The $N$ here denotes the rank of the gauge
group: an example of such theory, with gauge group $SU(N)$,
is discussed in section \ref{sec:QFT}. The large $N$ limit
consists of taking $N \to \infty$ while keeping fixed the 't Hooft
coupling constant $\lambda = g_{YM}^2 N$ \cite{'tHooft:1973jz}. One can show that in this limit
only planar diagrams survive. On the bulk side, taking the large $N$
limit means that one suppresses loop effects. The value of $\lambda$
then controls whether the supergravity approximation is valid or not.

The duality relates bulk fields with gauge-invariant operators of the
boundary theory. For example, the bulk metric corresponds to the boundary
stress-energy tensor, $T_{ij}$, while bulk scalar fields correspond to
boundary scalar operators, such as $\tr F_{ij} F^{ij}$, where $F_{ij}$ is
the field strength of the gauge field and the trace is over the gauge group indices. Correlation functions
of such gauge invariant operators can then be extracted from the
asymptotics of bulk solutions and conversely, given
correlation functions of the dual operators, one can reconstruct
asymptotic solutions.

To understand the holographic computations we need to know a few things
about the structure of asymptotic solutions of the field equations.
The discuss below refers specifically to a four dimensional bulk spacetime
(the general features are the same in any dimension but the details are different).
The results for the case of asymptotically power-law spacetimes can be obtained from the results for asymptotically $AdS_{2 \sigma+1}$ spacetimes via a dimensional reduction on a $T^{2 \sigma -3}$ torus and analytic continuation in $\sigma$ \cite{Kanitscheider:2009as}.
The most general asymptotic solution for both cases can be shown to be of the
form \cite{de Haro:2000xn, Kanitscheider:2008kd}
\begin{eqnarray}
\label{FG}
\d s^2 &=& \d r^2 + g_{ij}(r, x) \d x^i \d x^j, \label{reconstruction} \nn \\
g_{ij}(r, x) &=& e^{2r} \(g_{(0)ij}(x) + e^{- 2 r} g_{(2)ij}(x) + \cdots + e^{- 2 \sigma r} g_{(2 \sigma)ij}(x) + \cdots\),
\end{eqnarray}
where $\sigma=3/2$ for solutions of Einstein's equations with negative cosmological constant and
$\sigma = (3 n-1)/2 (n-1)>3/2$ for the case of asymptotically power-law solutions. In the latter case,
the metric above  is given in the so-called dual frame \cite{Boonstra:1998mp}:
the Einstein frame metric $g^E_{ij}$ in (\ref{Action}) is related to the metric $g^D_{ij}$ in (\ref{reconstruction})
via the Weyl transformation $g^E_{ij} = \exp(\lambda\Phi) g^D_{ij}$, where $\lambda=\sqrt{2/n}$.
In  (\ref{reconstruction}), $g_{(0)ij}(x)$ is an arbitrary (non-degenerate) three-dimensional metric, from which the $g_{(2 k)ij}(x)$ with $k<\sigma$ are locally determined, while $g_{(2 \sigma)ij}(x)$ is only partially constrained by the asymptotic analysis of the field equations.
One can show, however, that this coefficient is directly related
to the expectation value of the boundary stress-energy tensor \cite{de Haro:2000xn,Kanitscheider:2008kd}:
\[ \label{vev}
\langle T_{ij} \rangle =\frac{1}{2 \bar{\kappa}^2}\, (2 \sigma g_{(2 \sigma)ij}).
\]

Consider now a scalar field\footnote{In the
asymptotically AdS case, $\Psi=\Phi$, while in the asymptotically power-law case,
$\Psi = \exp ((n-1) \Phi/\sqrt{2n})$. To match with the discussion 
in \cite{Kanitscheider:2008kd} for the D$2$-brane case ($n=7$), note that 
$\Phi_{\rm here}= -(\sqrt{14}/5) \phi_{{\rm there}}$, 
where $\phi_{{\rm there}}$ is the scalar field in 
 \cite{Kanitscheider:2008kd} for $p=2$.}
$\Psi$ that is dual to an operator ${\cal O}$ of dimension $\Delta$.
The conformal dimension depends on the mass of $\Psi$ via $m^2 = \Delta (3 -\Delta)$ in the asymptotically AdS
case and is equal to $\Delta=4$ in the asymptotically power-law case (for appropriately normalized
${\cal O}$, see \cite{Kanitscheider:2008kd}).
The asymptotic expansion for $\Psi$ has a form analogous to (\ref{reconstruction}),
\[ \label{phivev}
\Psi(x,r) = e^{(\Delta-3) r} (\Psi_{(0)} + e^{-2 r} \Psi_{(2)} + \cdots
+ e^{- 2 \tilde{\sigma} r} \Psi_{(2 \tilde{\sigma})}(x) + \cdots),
\]
Here $\Psi_{(0)}(x)$ is unconstrained and is the source for the dual operator ${\cal O}$, with all subleading
terms, up to order $\exp (-2 \tilde{\sigma} r)$, then being locally determined in terms of the sources. The following term in the series, $\Psi_{(2 \tilde{\sigma})}$, is undetermined and is related to the expectation value of dual operator $\langle {\cal O} \rangle$. In the asymptotically AdS case,
$2 \tilde{\sigma} = (2 \Delta -3)$, while in the asymptotically power-law case, $\tilde{\sigma}=\sigma=(3 n-1)/2 (n-1)$.
(For details, see \cite{Skenderis:2002wp, Kanitscheider:2008kd}).

The constraints on $g_{(2 \sigma)ij}$ due to the Einstein equations imply
\begin{align} \label{WI}
& \nabla^i\<T_{ij}\>+\<\mathcal{O}\>\nabla_i\Psi_{(0)}=0, \\
& \<T^i_i\>+(3-\Delta)\Psi_{(0)} \<\mathcal{O}\>=0, \nonumber
\end{align}
and these are precisely the expected Ward identities\footnote{Note there is no conformal anomaly in three dimensions.}.

The relation (\ref{reconstruction})-(\ref{vev}) can be read in two ways: (i) given a supergravity
solution it allows us to read off the QFT data encoded by the solution; (ii) given QFT data
it provides a reconstruction of the bulk spacetime in the neighborhood of the boundary. Note that
the latter is true even when the supergravity approximation is not valid
in the interior of spacetime (because the curvature is large there). The terms exhibited in (\ref{reconstruction}), apart from $g_{(2 \sigma)ij}$,
are non-normalizable terms and are not affected by dynamics. The first term that is affected
by dynamics is $g_{(2 \sigma)ij}$ and this is indeed unconstrained asymptotically, except for the constraints
due to symmetries (namely the Ward identities (\ref{WI})). The gauge/gravity duality provides
a dual description of dynamics, so the statement of the duality is that $g_{(2 \sigma)ij}$ determined
from QFT via
(\ref{vev}) should agree with the value obtained from string dynamics in a spacetime
with these asymptotics. When the
gravity approximation is valid throughout, the asymptotics yield sufficient information
to uniquely reconstruct a regular bulk solution: the pair $(g_{(0)ij},\, g_{(2 \sigma)ij})$
(and similar
for the scalar field) are coordinates in the covariant phase space of the gravitational theory
\cite{Papadimitriou:2005ii}.

An alternative way to express these results that we will use in the next subsection
is to use the radial Hamiltonian formulation of \cite{Papadimitriou:2004rz}.
This is a Hamiltonian formulation where the radial direction plays the role of time.
In this formalism the expectation value of the stress-energy tensor is given by
\[
\label{radial_formula}
\<T_{ij}\> = \(\frac{-2}{\sqrt{g}}\,\bar{\Pi}_{ij}\)_{(3)},
\]
where $\bar{\Pi}_{ij}$
is the radial canonical momentum and we will momentarily explain the meaning of subscript $(3)$.
There is a similar formula for the expectation value of ${\cal O}$ that we will not need here.
A fundamental property of the
spacetimes (\ref{FG}) is that the radial derivative is, to leading order as $r \to \infty$,
proportional to the dilatation operator $\delta_D$,
\[ \label{r-RG}
\partial_r = \delta_D (1 + O[e^{-2 r}]),
\]
where
\[ \label{dilat}
\delta_D g_{ij}(x,r) = 2 g_{ij}(x,r), \qquad \delta_D \Psi(x,r) = (\Delta -3) \Psi(x,r)
\]
(one can verify (\ref{r-RG}) by inspection of (\ref{FG}) and (\ref{phivev})). The expressions in (\ref{dilat})
are the standard dilatation transformation rules for a metric and a source that couples to
an operator of dimension $\Delta$ in three spacetime dimensions. Equation (\ref{r-RG})
is a precise version of the
often quoted relation between the radial direction and the energy scale of the dual QFT.
In our context, equation (\ref{r-RG}) implies that one can trade the
radial expansion for an expansion in terms of eigenfunctions of dilatation operator.
An eigenfunction $A_{(n)}$ of weight $n$
is by definition,
\[
\delta_D A_{(n)} = - n A_{(n)}.
\]
As follows from (\ref{r-RG}),  $A_{(n)} \sim e^{-n r} (1 + O[e^{-2 r}])$,
so the radial expansion and the expansion is eigenfunctions of the dilatation operator
are closely related.
The chief advantage of the radial Hamiltonian reformulation is that the expansion is now manifestly covariant, whereas
expanding in a particular coordinate is not a covariant operation.
Now, the radial canonical momentum can be decomposed in eigenfunctions of the dilatation
operator and the subscript $(3)$ in (\ref{radial_formula}) indicates that we should
pick the part with
dilatation eigenvalue\footnote{In odd (bulk) dimensions, the transformation rule of
this specific coefficient has
also an additional anomalous contribution
due to the conformal anomaly. There is no conformal
anomaly in our case, and this coefficient is a true eigenfunction of the dilatation operator.}
$3$. One may have expected this on general grounds, since the dimension of stress energy tensor is
three dimensions is $3$. More generally, the expectation value of an operator of dimension $\Delta$
is given by the piece of the corresponding radial canonical momentum of weight $\Delta$.
Note that in general the radial canonical momentum contains also pieces with weight less than $\Delta$
({\it i.e.}, less than $3$ in the case of $T_{ij}$): these pieces diverge as $r \to \infty$. One of the advantages  of radial
Hamiltonian formalism \cite{Papadimitriou:2004rz} is that renormalization is simple: one simply discards all pieces with weight less than $\Delta$ (plus additional logarithmic divergences at weight $\Delta$, when present). One can indeed show that removing these pieces is equivalent to adding local covariant
counterterms to the on-shell action.

The radial canonical momentum for the asymptotically AdS case is equal to
\[
\label{pi_EF}
\bar{\Pi}_{ij}=\frac{1}{2\tK^2}\sqrt{g}(K_{ij}-Kg_{ij}),
\]
with $K_{ij}=(1/2)\partial_r g_{ij}$ the extrinsic curvature of constant-$r$ slices.
In the case of asymptotically power-law backgrounds, the relevant canonical momentum is that of the dual frame \cite{Kanitscheider:2008kd}, namely
\[
\label{pi_DF}
\bar{\Pi}^{D}_{ij} = \frac{1}{2\tK^2}\sqrt{g}e^{\lambda\Phi}\(K_{ij}-(K+\lambda\partial_r{\Phi})\delta^i_j\),
\]
where $\lambda=\sqrt{2/n}$ and the metric and extrinsic curvature are those of the dual frame.

\subsection{2-point functions.}

To linear order in the sources, the variation of the 1-point function for the stress-energy tensor is
\[
\delta \<T_{ij}(x)\> = -\int\d^3 y \sqrt{g_{(0)}}\( \frac{1}{2}\<T_{ij}(x)T_{kl}(y)\>\delta g^{kl}_{(0)}(y)+\<T_{ij}(x){\cal{O}}(y)\>\delta\vphi_{(0)}(y)\).
\]
It follows that in order to obtain the holographic $2$-point functions we need to solve the
the linearized equations of motion about the domain-wall solution with Dirichlet boundary conditions at infinity and imposing regularity in the interior.
Transforming to momentum space, on general grounds, the 2-point function takes the form
\[
\label{ABdef}
\<T_{ij}(\tq)T_{kl}(-\tq)\> = A(\tq)\Pi_{ijkl}+B(\tq)\pi_{ij}\pi_{kl},
\]
where $\Pi_{ijkl}$ is the three-dimensional transverse traceless projection operator defined by
\[
\Pi_{ijkl}= \frac{1}{2}(\pi_{ik}\pi_{lj}+\pi_{il}\pi_{kj}- \pi_{ij}\pi_{kl}), \qquad
\pi_{ij} = \delta_{ij} - \frac{\bar{q}_i \bar{q}_j}{\bar{q}^2}.
\]
Decomposing the boundary metric as in (\ref{metric_decomp}), we then find
\[
\label{ref_formula}
\delta\<T_{ij}\> = \frac{1}{2}A(\tq)\g_{(0)ij}-2B(\tq)\psi_{(0)}\pi_{ij}-\<T_{ij}(\tq){\cal{O}}(-\tq)\>\delta\vphi_{(0)}.
\]
From (\ref{radial_formula}), this expression is to be compared with the bulk radial canonical momentum expanded to linear order.
We begin by writing the perturbed metric as in (\ref{lin_fl}) and (\ref{metric_decomp}), with the lapse and shift perturbations gauged to zero ($\phi=\nu=\nu_i=0$), and we additionally set $\omega_i$ to zero using the constraints (\ref{constraints}) and a spatial gauge transformation.
Then, for the case of asymptotically AdS backgrounds, using (\ref{pi_EF}) we find 
\[
\delta\<T^i_j\> = \frac{1}{\tK^{2}}[\delta K\delta^i_j-\delta K^i_j]_{(3)}=
-\frac{1}{\tK^{2}}\left[2\dot{\psi}\delta^i_j+\tq^2\dot{\chi}\pi^i_j+\frac{1}{2}\dot{\gamma}^i_j\right]_{(3)}.
\]
Expressing (\ref{constraints}) in the present gauge, it follows that
\begin{align}
2\dot{\psi}&=\dot{\vphi}\delta\vphi, \\
\tq^2 \dot{\chi}&=\frac{\tq^2\psi}{a^2H}-\ep\dot{\zeta}
=\frac{\tq^2\psi}{a^2H}+\frac{\tK^2\bar{\Omega}\zeta}{2a^3}
=\(\frac{\tq^2}{a^2H}+\frac{\tK^2\bar{\Omega}}{2a^3}\)\psi+\frac{\tK^2\bar{\Omega}H}{2a^3\dot{\vphi}}\delta\vphi,
\nonumber
\end{align}
where in the last line we have used the definition of the domain-wall response function $\bar{\Omega}$ in (\ref{dw_response_fns}) and expanded $\zeta$ according to (\ref{zeta}).
Then, using the definition of the response function $\bar{E}$ also in (\ref{dw_response_fns}), we find
\[ \label{delta_pi}
\delta\<T^i_j\> = \left[\frac{2\bar{E}}{a^3}\gamma^i_j - \(\frac{\tq^2}{\tK^2 a^2 H}+\frac{\bar{\Omega}}{2a^3}\)\psi\pi^i_j-
\(\frac{H\bar{\Omega}}{2a^3\dot{\vphi}}\pi^i_j+\frac{\dot{\vphi}}{\tK^2}\delta^i_j\)\delta\vphi\right]_{(3)}.
\]
Since  the scale factor $a$ has dilatation weight $-1$ (as follows from (\ref{dilat})), comparison with (\ref{ref_formula}) yields\footnote{The term proportional to $\tq^2$ in the coefficient of $\psi$ in (\ref{delta_pi}) only contributes a contact term to $B(\tq)$ which we drop.}
\[
\label{ABeqns}
A(\tq) =  4 \bar{E}_{(0)}(\tq) , \qquad
B(\tq) = \frac{1}{4} \bar{\Omega}_{(0)}(\tq),
\]
where the zero subscript indicates the pieces of the response
functions that have zero weight under dilatations, so in particular
they are independent of $r$ as $r \to \infty$.  Note that, in general,
$\bar{E}$ and $\bar{\Omega}$ diverge as $r \to \inf$. 
Extracting $\bar{E}_{(0)}$ and
$\bar{\Omega}_{(0)}$ correctly then requires first determining the terms with
eigenvalue less than zero and subtracting these from $\bar{E}$ and
$\bar{\Omega}$, before taking the limit $r \to \infty$ (see
\cite{Papadimitriou:2004rz} and the example in the next subsection).
The issue here is that the subtraction of the infinite pieces may 
induce a change in the finite part as well. This can happen if the 
local covariant counterterms needed to cancel the infinities
necessarily have a finite part as well.

Considering now the case of backgrounds that are asymptotically power-law, as before we decompose the metric perturbations as in (\ref{lin_fl}) and (\ref{metric_decomp}), and gauge the shift and vector perturbations to zero ($\nu=\nu_i=\omega_i=0$). This time, however, we choose the lapse perturbation to be $\phi=(\lambda/2)\delta\vphi$.
Transforming to the dual frame, we then find
\begin{align}
\d\tilde{s}^2 &= e^{-\lambda\Phi}\d s^2 = \d r^2 + \tilde{a}^2[\delta_{ij}+\tilde{h}_{ij}]\d x^i \d x^j,\nn \\
\tilde{h}_{ij} &= -2\tilde{\psi}\delta_{ij}+2\partial_i\partial_j\tilde{\chi}+\tilde{\g}_{ij},
\end{align}
where $\tilde{a} = \exp(-\lambda\vphi/2)a$, with $\lambda{=}\sqrt{2/n}$ and the radial coordinate $\d r = \exp(-\lambda\vphi/2)\d z$.  The dual frame perturbations are related to their Einstein frame counterparts by
\[
\tilde{\psi}=\psi+(\lambda/2)\delta\vphi, \qquad \tilde{\chi}=\chi, \qquad \tilde{\g}_{ij}=\g_{ij}.
\]
From (\ref{radial_formula}) and (\ref{pi_DF}), we have
\begin{align}
\delta\<T^i_j\> &= \frac{1}{\tK^{2}}\left[e^{\lambda\vphi}\((\delta \tilde{K}+\lambda\delta\vphi_{,r})\delta^i_j-\delta\tilde{K}^i_j +(\ldots)\delta\vphi \) \right]_{(3)} \nn \\
&=-\frac{1}{\tK^{2}}\left[ e^{\lambda\vphi} \( 2\tilde{\psi}_{,r}\delta^i_j-\lambda\delta\vphi_{,r}\delta^i_j+\tq^2\tilde{\chi}_{,r}\pi^i_j+\frac{1}{2}\tilde{\gamma}^i_{j,r}+(\ldots)\delta\vphi\)\right]_{(3)} \nn \\
&= -\frac{1}{\tK^{2}}\left[ e^{3\lambda\vphi/2} \(2\dot{\psi}\delta^i_j+\tq^2\dot{\chi}\pi^i_j+\frac{1}{2}\dot{\gamma}^i_j+(\ldots)\delta\vphi\)\right]_{(3)}.
\end{align}
It follows from (\ref{ref_formula}) that the terms proportional to $\delta\vphi$ do not contribute to the stress tensor 2-point function that we are interested in, so in the above (and below) these terms have been suppressed.
The Einstein frame constraints (\ref{constraints}), when expressed in the present gauge, yield
\[
2\dot{\psi} = (\ldots)\delta\vphi, 
\qquad
\tq^2 \dot{\chi} = \(\frac{\tq^2}{a^2H}+\frac{\tK^2\bar{\Omega}}{2a^3}\)\psi+(\ldots)\delta\vphi.
\]
Thus we have
\[
\delta\<T^i_j\>=\left[\frac{2\bar{E}}{\tilde{a}^3}\tilde{\gamma}^i_j -\(\frac{\tq^2e^{\lambda\vphi/2}}{\tK^2\tilde{a}^2H}+\frac{\bar{\Omega}}{2\tilde{a}^3}\)\tilde{\psi}\pi^i_j+(\ldots)\delta\vphi \right]_{(3)},
\]
and since the dilatation weight of $\tilde{a}$ is $-1$, we again recover (\ref{ABeqns}) modulo contact terms.

\subsection{Example: Power-law inflation.}

To illustrate the above discussion, let us consider the domain-wall
 backgrounds equal (rather than asymptotic) to (\ref{power_law}) discussed in  \cite{Kanitscheider:2008kd}, namely
\[
a=(z/z_0)^n, 
\qquad \vphi=\sqrt{2n}\ln(z/z_0), 
\qquad z_0 = n-1 > 0.
\]
Under the domain-wall/cosmology correspondence these solutions are mapped to cosmologies undergoing exact power-law inflation. While this model is strongly constrained by the WMAP data \cite{Komatsu:2008hk}, this need not concern us here since our purpose is simply to illustrate the steps involved in the holographic computation. Furthermore, in Section \ref{sec:holo_pheno} we will see that the strong coupling version of these
models ({\it i.e.}, where gravity is strongly coupled at early times but the dual three-dimensional QFT is
weakly coupled) are compatible with observations.

Following \cite{Skenderis:2006jq}, one can obtain the fake superpotential from the solution
yielding $W = -(2n/z_0) \exp(-\varphi/\sqrt{2n})$. It follows that $\ep =1/n$ and both $\g_{ij}$ and $\zeta$ obey the same equation of motion, which for the domain-wall spacetime reads
\[ \label{zetaeqn}
0=\ddot{\z}_{\bar{q}}+(3n/z)\dot{\z}_{\bar{q}}-(z/z_0)^{-2n}\bar{q}^2 \z_{\bar{q}}.
\]
Imposing regularity in the interior, the solution is
\[
\z_{\bar{q}} = C_{\bar{q}} \rho^\sigma K_{\sigma}(\rho),
\]
where $K_\sigma$ is a modified Bessel function of the second kind of order $\sigma = (3n-1)/2 (n-1)>3/2$, $C_{\bar{q}}$ is an arbitrary function of $\bar{q}$ and the radial coordinate $\rho=\bar{q} (z/z_0)^{1-n}$.
The boundary $z\tto \inf$ corresponds to $\rho = 0$ while the domain-wall interior corresponds to $\rho \tto \inf$. The corresponding radial canonical momentum is equal to
\[ \label{can_mom}
\bar{\Pi}_{\bar{q}}^{(\z)}=\frac{2 \ep}{\bar{\kappa}^2} a^3 \dot{\zeta}_{\bar{q}} =-\frac{2C_{\bar{q}}}{n\bar{\kappa}^2}\(\frac{\rho}{\bar{q}}\)^{-2\sigma} \rho\partial_\rho (\rho^\sigma K_\sigma (\rho)).
\]
Expanding about $\rho=0$, we find
\begin{align}
\z_{\bar{q}} &= C_{\bar{q}} \( 1+\frac{1}{4(1-\sigma)}\rho^2 +\ldots -\frac{\Gamma(1-\sigma)}{4^\sigma \Gamma(1+\sigma)}\rho^{2\sigma}+\ldots\), \nn \\
\bar{\Pi}^{(\z)}_{\bar{q}} &= -C_{\bar{q}}\frac{2\bar{q}^{2\sigma}}{n\bar{\kappa}^2}  \( \frac{1}{2(1-\sigma)}\rho^{2(1-\sigma)} +\ldots -\frac{2\sigma\Gamma(1-\sigma)}{4^\sigma \Gamma(1+\sigma)}+\ldots\),
\end{align}
and thus
\[ \label{Omega}
\Omega(\bar{q})=-\frac{\bar{\Pi}^{(\z)}_{\bar{q}}}{\z_{\bar{q}}} =
\frac{2\bar{q}^{2\sigma}}{n\bar{\kappa}^2}  \( \frac{1}{2(1-\sigma)}\rho^{2(1-\sigma)} +\ldots -\frac{2\sigma\Gamma(1-\sigma)}{4^\sigma \Gamma(1+\sigma)}+\ldots\).
\]
As expected, this diverges as $\rho \to 0$. To compute the 2-point function we need to identify the
parts that have negative dilatation eigenvalue, subtract them from (\ref{Omega}) and then
take $\rho \to 0$.

To do this, we first transform to the dual frame via $g^D_{ij}=e^{-\sqrt{(2/n)}\vphi}g^E_{ij}$
and then change radial variable, $r = z_0 \ln (z/z_0)=-\ln(\rho/\tq)$. The metric is now that of AdS,
\[
\d s^2 = \d r^2 + e^{2 r} \d\vec{x}^2,
\]
and the dilatation operator is exactly equal to the radial derivative,
\[
\delta_D = \partial_r = - \rho \partial_\rho.
\]
(This reflects the fact that the AdS isometry group is the same as the conformal group in
one dimension less). It follows that any monomial in $\rho$ is an eigenfunction of
$\delta_D$,
\[
\delta_D \rho^n = - n \rho^n,
\]
and one can simply identify in (\ref{Omega}) all terms with negative eigenvalue; for example, $\bar{\Omega}_{(-2\sigma+2)}= \bar{q}^{2\sigma}/(n\bar{\kappa}^2 (1{-}\sigma))\rho^{-2(\sigma-1)}$.
We then have
\[
\label{Omega0}
\bar{\Omega}_{(0)}=-\frac{4\sigma \Gamma(1-\sigma)}{n 4^\sigma \Gamma(1+\sigma)}\bar{\kappa}^{-2}\bar{q}^{2\sigma}.
\]

In this example, the identification of the terms with negative eigenvalues could be accomplished by inspection.
In more complicated examples, however, this is no longer the case, so we briefly indicate here how one could
compute them (see \cite{Papadimitriou:2004rz} for a more complete discussion).
Starting from (\ref{zetaeqn}), and using the definition of the corresponding
canonical momentum (the first equality in (\ref{can_mom})), one obtains
\[
\dot{\bar{\Pi}}_{\bar{q}} - \frac{2 \bar{q}^2}{n \bar{\kappa}^2} a \zeta_{\bar{q}}=0.
\]
Inserting in this equation the definition of the response function, and changing the radial
coordinate from $z$ to $r$, one obtains
\[ \label{eqnOmega}
\partial_r \bar{\Omega} - \alpha^2 \bar{\Omega}^2 e^{-2 \sigma r} + (\bar{q}^2/\alpha^2) e^{2 (\sigma-1)r} =0,
\]
where $\alpha^2 = n \bar{\kappa}^2/2$. This may now be solved asymptotically by expanding $\Omega$
in dilatation eigenvalues,
\[
\bar{\Omega} = \sum_{k \geq 1} \bar{\Omega}_{(-2 \sigma +2 k)},
\]
making use\footnote{In examples where the background solution is only asymptotically AdS, the relation between the dilatation operator and the radial derivative 
contains subleading terms (see (\ref{r-RG})) and these 
that must be taken into account
in this computation, see \cite{Papadimitriou:2004rz} for a complete discussion.} of
 $\partial_r = \delta_D$ and collecting all terms with the same weight. For example, to
leading order, at weight $(-2 \sigma {+} 2)$, only the first and last term in (\ref{eqnOmega})
can have this weight, and one obtains
$
\bar{\Omega}_{(-2 \sigma+2)} = (\bar{q}^2/\alpha^2 2(1{-}\sigma)) \exp (2 (\sigma{-}1)r)
$
in agreement with (\ref{Omega0}). Through iteration, one may obtain all coefficients with negative eigenvalue.

Having obtained $\bar{\Omega}_{(0)}$, we may finally compute $B(\bar{q})$:
\[
B(\bar{q}) = \frac{1}{4}\bar{\Omega}_{(0)}= -\frac{\sigma \Gamma(1-\sigma)}{n 4^\sigma \Gamma(1+\sigma)}\bar{\kappa}^{-2}\bar{q}^{2\sigma} = -\frac{\pi }{ 4^\sigma\Gamma^2(\sigma) \,n\sin\pi\sigma}\,\tK^{-2}\tq^{2\sigma}.
\]
A near-identical argument holds for the tensors $\g_{ij}$ yielding $\bar{\Omega}_{(0)} = (8/n) \bar{E}_{(0)}$, and hence $A(\bar{q}) = 2n B(\bar{q})$.
Via the domain-wall/cosmology correspondence, applying the continuations (\ref{a/c}) to (\ref{Omega0}), the imaginary parts of the cosmological response functions are
\[
\label{imom}
\Im \Omega_{(0)} =(8/n)\Im E_{(0)}= -\frac{4\pi}{n 4^\sigma \Gamma^2(\sigma)}\kappa^{-2}q^{2\sigma}.
\]
From (\ref{cosmo_result}), we then recover the expected cosmological power spectra:
\[
\label{powerlawspectrum}
\Delta^2_S(q) =\frac{n}{16} \Delta^2_T(q)= \frac{n 4^{\sigma-2}\Gamma^2(\sigma)}{\pi^3}\, \K^2 q^{3-2\sigma}.
\]

Note that we could equally well have obtained (\ref{imom}) by applying the continuations (\ref{a/c}) to the {\it unrenormalized} domain-wall response function (\ref{Omega}), and then taking the imaginary part followed by the limit $z\tto \inf$.  This is because the divergent terms one subtracts to obtain the renormalized reponse functions are all analytic functions of $\tq^2$ (as may be seen from (\ref{Omega}), where the leading term is proportional to $\tq^2$) and hence under the continuation $\tq^2=-q^2$, these terms remain real and do not contribute to the imaginary part of the cosmological response functions.  Only the leading {\it non-analytic} piece of the domain-wall response functions contributes to the late-time imaginary part of the cosmological response functions: this leading non-analytic piece is finite and is simply $\bar{\Omega}_{(0)}$.
In fact, the late-time values of the imaginary parts of the cosmological response functions have to be finite as a consequence the Wronskian relations (\ref{Wronsk}) and the fact that $\zeta$ and $\g_{ij}$ tend to a finite constant at late times.

\section{Continuation to the pseudo-QFT.} \label{sec:anal_cont}

We now wish to re-express the bulk analytic continuation (\ref{a/c}) in terms of QFT variables,
corresponding to the vertical line on the right-hand side of Fig.~1.  First, the dimensionful coupling of the theory, corresponding
to the deformation by the operator ${\cal O}$ can be read off from the asymptotics of $\Phi$.
Since $\Phi$ does not continue, neither does the coupling constant.
Second, $\bar{\kappa}^{-2}$ is proportional to the square of the number of colors, $\bar{N}^2$.
It follows that the continuation (\ref{a/c}) amounts to
\[
\bar{N}^2 = - N^2,  \qquad \tq = -i q,
\]
where the barred quantities are associated with the QFT dual to the domain-wall and the unbarred quantities are associated with the pseudo-QFT dual to the cosmology.  (Thus $\bar{N}$ is the rank of the gauge group of the QFT dual to the domain-wall spacetime, while $N$ is the rank of gauge group of the pseudo-QFT dual to the cosmological spacetime).
We therefore find that the power spectrum for any inflationary cosmology that is
asymptotically de Sitter or asymptotically power-law can be
directly computed from the 2-point function of a three-dimensional QFT
via the formulae:
\[
\label{result}
\Delta^2_S(q) = \frac{-q^3}{16 \pi^2 \Im B(-i q)}, \qquad
\Delta^2_T(q) = \frac{-2 q^3}{\pi^2 \Im A(-i q)}.
\]
This is one of our principal results.

\section{Beyond the weak gravitational description.} \label{sec:beyond}

\begin{figure}[t]
\center
\includegraphics[width=7cm]{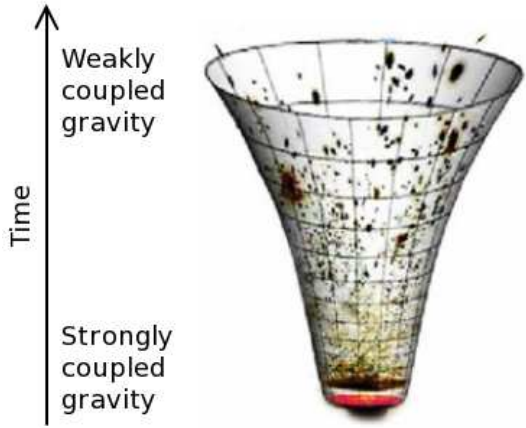}
\hspace{2pc}
\begin{minipage}[b]{14pc}
\caption{\label{univ_graphic}
Using holography it is possible to describe the generation of primordial cosmological perturbations during an early time epoch in which the gravitational description is {\it strongly coupled}.  At later times we envision a smooth transition to a conventional hot big bang description in which gravity is weakly coupled.  (Figure adapted from \cite{xfig}).

}
\end{minipage}
\end{figure}

In the discussion so far we have assumed that the description in terms
of gravity coupled to a scalar field is valid at early times, and that
the perturbative quantization of fluctuations can be justified.  The
holographic description also allows us to obtain results when these
assumptions do not hold.  At early times, the theory may be strongly
coupled with no useful description in terms of low-energy fields (such
as the metric and the scalar field).  The holographic set-up allows us
to extract the late-time behavior of the system, which can be
expressed in terms of low-energy fields, from QFT correlators.
This is the counterpart of the discussion in Section \ref{sec:Hol_an}
where we saw that in gauge/gravity duality
the asymptotic behavior of bulk fields near the boundary of
spacetime is reconstructed by the correlators of the dual QFT.
This late-time behavior is
precisely the information we need to compute the primordial power
spectra and other cosmological observables.  We will assume that this postulated early-time phase in which gravity is strongly coupled is subsequently followed by a smooth transition to the usual hot big bang cosmology in which gravity is weakly coupled, see Fig.~\ref{univ_graphic}.

\section{Holographic phenomenology for cosmology.} \label{sec:holo_pheno}

Ideally one would deduce from a string/M-theoretic construction what
the dual QFT is. Instead we initiate here a holographic
phenomenological approach. The dual QFT would involve scalars,
fermions and gauge fields and it should admit a large $N$ limit. The
question is then whether one can find a theory which is compatible
with current observations.  In particular, one might consider either
deformations of CFTs or theories with a single dimensionful parameter in the
regime where the dimensionality of the coupling constant drives the dynamics,
as these QFTs have already featured in our discussion above.

\subsection{A prototype dual QFT.} \label{sec:QFT}

We will discuss here super-renormalizable theories that contain one
dimensionful coupling constant.  A prototype example is three-dimensional $SU(\bar{N})$
Yang-Mills theory\footnote{We write the rank of the QFT gauge group here as $\bar{N}$ since we will first be performing calculations using the QFT dual to the domain-wall spacetime before analytically continuing to the pseudo-QFT.} coupled to a number of scalars and fermions, all transforming in the adjoint of $SU(\bar{N})$.
Theories of this type are typical in AdS/CFT
where they appear as the worldvolume theories of D-branes.
A general such model that admits a large $\bar{N}$ limit is
\begin{align}
\label{Lfree}
S &= \frac{1}{g_{\mathrm{YM}}^2}\int \d^3 x
\tr \Big[ \frac{1}{2} F^I_{ij}F^{Iij} +  \frac{1}{2} (D\phi^J)^2
+  \frac{1}{2} (D\chi^K)^2
+ \bar{\psi}^L \slashed{D} \psi^L \nn \\
&\qquad\qquad\qquad
+ \lambda_{M_1 M_2 M_3 M_4} \Phi^{M_1} \Phi^{M_2} \Phi^{M_3} \Phi^{M_4}
+ \mu_{M L_1 L_2}^{\alpha \beta} \Phi^M \psi^{L_1}_\alpha \psi^{L_2}_\beta \Big],
\end{align}
where we consider $\mathcal{N}_A$ gauge fields $A^I \ (I=1,\ \ldots,\ \mathcal{N}_A)$;
$\mathcal{N}_\phi$ minimal scalars $\phi^J\ (J=1,\ \ldots,\ \mathcal{N}_\phi)$;
$\mathcal{N}_\chi$ conformal scalars $\chi^K\ (K=1,\ \ldots,\ \mathcal{N}_\chi)$
 and $\mathcal{N}_\psi$ fermions $\psi^L \ (L=1,\ \ldots,\  \mathcal{N}_\psi)$.
The couplings $\lambda_{M_1 M_2 M_3 M_4}$ and
 $\mu_{M L_1 L_2}^{\alpha \beta}$ (where $\alpha$ and $\beta$ are spinor
indices) are dimensionless, and we have grouped the scalars appearing in the interaction terms as
$\Phi^M = (\{\phi^J\},\{\chi^K\})$. Note that all terms in the Lagrangian have dimension 4
so this QFT is indeed of the same type as the QFTs dual to asymptotically power law
solutions. The  operator ${\cal O}$ that featured in our earlier discussion of holography 
is closely related to the Lagrangian (see the discussion
in Section 4 of \cite{Kanitscheider:2008kd}).

The conformally coupled scalars have an $R \chi^2$ coupling when
we couple the theory to gravity; on a flat background this means the conformally coupled scalars have a different stress energy tensor from their minimally coupled counterparts.
Specifically, on a flat background, the stress tensor is given by
\begin{align}
 T_{ij} &= \frac{1}{g_{\mathrm{YM}}^2} \tr \Big[2F^I_{ik}F^{I\,k}_j
+D_i\phi^J D_j\phi^J+D_i\chi^K D_j\chi^K-\frac{1}{8}D_iD_j(\chi^K)^2
+\frac{1}{2}\bar{\psi}^L\g_{(i}{\overleftrightarrow{D}}_{j)}\psi^L  \nn \\
&\qquad\qquad -\delta_{ij}\Big( \frac{1}{2}F^I_{kl}F^{I\, kl}+\frac{1}{2}(D\phi^J)^2 +\frac{1}{2}(D\chi^K)^2-\frac{1}{8}D^2(\chi^K)^2 \nn \\
& \qquad\qquad \qquad\qquad
+\lambda_{M_1 M_2 M_3 M_4}\Phi^{M_1} \Phi^{M_2} \Phi^{M_3} \Phi^{M_4} + \mu_{M L_1 L_2}^{\alpha \beta}\Phi^M \psi^{L_1}_\alpha \psi^{L_2}_\beta\Big)\Big].
\end{align}

To extract predictions, we need to compute the
coefficients $A(\bar{q})$ and $B(\bar{q})$ appearing in the
general decomposition (\ref{ABdef}) of the stress tensor 2-point function, analytically continue the
results, and then insert them in the holographic formulae (\ref{result}) for the power
spectra.
To facilitate comparison with the observational data, we will additionally make use of the standard cosmological 
parameterizations 
\[
\label{defs}
 \Delta^2_S(q) = \Delta^2_S(q_0) \(\frac{q}{q_0}\)^{n_S(q)-1}, \qquad \Delta^2_T(q) = \Delta^2_T(q_0) \(\frac{q}{q_0}\)^{n_T(q)},
\]
where $\Delta^2_{S/T}(q_0)$ is the scalar/tensor amplitude at some chosen pivot scale $q_0$,
and $n_{S/T}(q)$ is the scalar/tensor spectral tilt.

\subsection{1-loop calculation.}

\begin{figure}[t]
\center
\includegraphics[width=5.8cm]{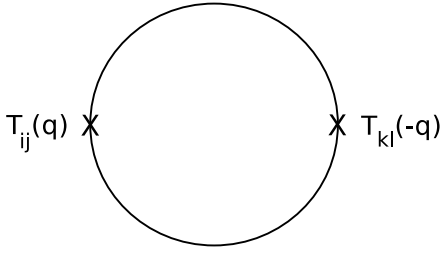} 
\hspace{2pc}
\begin{minipage}[b]{14pc}
\caption{\label{1-loop-dia}
1-loop contribution to $\<T_{ij}(\bar{q})T_{kl}(-\bar{q})\>$.  We sum over the contributions from gauge fields, scalars and fermions, with each diagram yielding a contribution of order $\sim \bar{N}^2\bar{q}^3$.}
\end{minipage}
\end{figure}

The leading contribution to the 2-point function of the stress tensor is at one loop (see Fig.~\ref{1-loop-dia}).
Since the stress tensor has dimension three, and the only dimensionful quantity that can appear to this order is $\bar{q}$ (1-loop amplitudes are independent of $g_{YM}^2$),
it follows that
\[
\label{1-loop_result}
A(\bar{q}) = C_A \bar{N}^2 \bar{q}^3 + O(g_{\mathrm{YM}}^2),
\qquad  B(\bar{q}) = C_B \bar{N}^2 \bar{q}^3 + O(g_{\mathrm{YM}}^2),
\]
where $C_A$ and $C_B$ are numerical coefficients whose value depends only on the field content.
Explicit calculation then reveals that
\[
C_A = (\mathcal{N}_A + \mathcal{N}_\phi +\mathcal{N}_\chi+2\mathcal{N}_\psi)/256, \qquad
C_B = (\mathcal{N}_A+\mathcal{N}_\phi)/256.
\]
Inserting (\ref{1-loop_result}) into our holographic formulae (\ref{result}),
we find
\[
\label{ampl-to-1-loop}
 \Delta_S^2(q) = \frac{1}{16\pi^2 N^2 C_B} +O(g_{\mathrm{YM}}^2), \qquad \Delta_T^2(q) = \frac{2}{\pi^2 N^2 C_A}+O(g_{\mathrm{YM}}^2).
\]
Comparing with (\ref{defs}), we immediately see that the power spectra are {\it scale-invariant} to leading order ({\it i.e.}~$n_S=1+O(g_\mathrm{YM}^2)$, $n_T=O(g_\mathrm{YM}^2)$), regardless of the precise field content of the model.
To estimate the value of $N$ we may compare with the observed
amplitude of the scalar power spectrum.  From the WMAP data \cite{Komatsu:2008hk} we have $\Delta^2_S(q_0) \sim O(10^{-9})$, hence $N \sim O(10^4)$, justifying our use of the large $N$ limit.

The observational data also serve to provide an upper bound on the ratio of tensor to scalar power spectra.
From (\ref{ampl-to-1-loop}), we find
\[
 r = \Delta_T^2 / \Delta_S^2 = 32 C_B / C_A,
\]
and hence an upper bound on $r$ translates into a constraint on the field content of the dual QFT.  A smaller upper bound on $r$ requires increasing the number of conformal scalars and massless fermions and/or decreasing the number of gauge fields and minimal scalars.

\subsection{2-loop corrections.}

\begin{figure}[t]
\center
\includegraphics[width=0.9\textwidth]{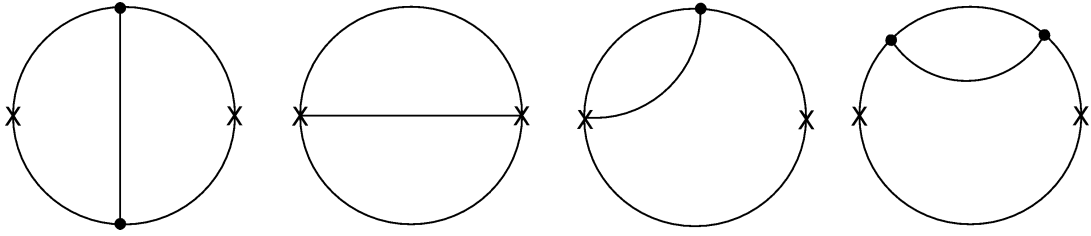} 
\caption{\label{2-loop-fig} Diagram topologies contributing at 2-loop order. 
Each diagram consists of an overall factor of
$\bar{N}^3 g_{\mathrm{YM}}^2$ multiplying an integral with superficial degree of divergence two.  After dimensional regularization and renormalization, the integrals evaluate to $\sim \bar{q}^2 \ln (\bar{q}/\bar{q}_*)$, and so overall each diagram yields a contribution to the stress tensor 2-point function of order $\sim \bar{N}^3 g_{\mathrm{YM}}^2\bar{q}^2 \ln (\bar{q}/\bar{q}_*)$, or equivalently $\sim \bar{N}^2\bar{q}^3 g_{\mathrm{eff}}^2 \ln (\bar{q}/\bar{q}_*)$.}
\end{figure}

Corrections to the stress tensor 2-point function at 2-loop order\footnote{Super-renormalizable theories have infrared divergences, but large $N$ resummation leads to well-defined expressions with $g_{\mathrm{YM}}^2$ effectively playing the role of an infrared regulator. The exact amplitudes are nonanalytic functions of the coupling constant  \cite{Jackiw:1980kv}.  Note that our analytic continuation to pseudo-QFT does not involve the coupling constant.}
 give rise to small deviations from scale invariance.
The full result will be reported elsewhere \cite{to_appear}, however, it is easy to obtain an order of magnitude estimate on general grounds.
The perturbative expansion depends on the effective dimensionless coupling constant $g_{\mathrm{eff}}^2 = g_{\mathrm{YM}}^2 \bar{N}/\tq$.  Either from inspection or from direct calculation of some of the diagrams contributing at $O(g_\mathrm{eff}^2)$ (see Fig.~\ref{2-loop-fig}), one finds
\begin{align}
\label{2-loop_result}
 A(\bar{q}) &= C_A \bar{N}^2 \bar{q}^3[1+D_A g_{\mathrm{eff}}^2 \ln (\bar{q}/\bar{q}_*)+O(g_{\mathrm{eff}}^4)], \nn \\
B(\bar{q}) &= C_B \bar{N}^2 \bar{q}^3[1+D_B g_{\mathrm{eff}}^2 \ln (\bar{q}/\bar{q}_*)+O(g_{\mathrm{eff}}^4)],
\end{align}
where $D_A$ and $D_B$ are numerical coefficients of order one whose value depends only on the field content.  To compute $D_A$ and $D_B$ precisely requires summing all the relevant 2-loop diagrams.

Inserting these two-loop corrected results into the holographic formulae (\ref{result}), we find
\begin{align}
\Delta_S^2(q) &= \frac{1}{16\pi^2 N^2 C_B}[1-D_B g_{\mathrm{eff}}^2\ln (q/q_*) +O(g_{\mathrm{eff}}^4)], \nn \\
\Delta_T^2(q) &= \frac{2}{\pi^2 N^2 C_A}[1-D_A g_{\mathrm{eff}}^2 \ln (q/q_*)+O(g_{\mathrm{eff}}^4)],
\end{align}
where the analytically continued effective coupling $g_{\mathrm{eff}}^2 = g_{\mathrm{YM}}^2 N/q$.
In comparison, expanding (\ref{defs}) yields
\begin{align}
\Delta_S^2(q) &= \Delta_S^2(q_0) [ 1+(n_S(q){-}1) \ln (q/q_0) + O\((n_S(q){-}1)^2\)], \nn \\
\Delta_T^2(q) &= \Delta_T^2(q_0) [1+n_T(q) \ln (q/q_0) + O\(n_T(q)^2\)].
\end{align}
Identifying the renormalization scale $q_*$ with the pivot scale $q_0$, we then see that the spectral amplitudes given in (\ref{ampl-to-1-loop}) are correct to $O(g_{\mathrm{eff}}^4)$, and that the corresponding spectral tilts are
\[
 n_S(q){-1} = -D_B g_{\mathrm{eff}}^2 + O(g_{\mathrm{eff}}^4), \qquad
n_T(q) = -D_A g_{\mathrm{eff}}^2 + O(g_{\mathrm{eff}}^4).
\]
Comparing with the WMAP data, from Table 4 of \cite{Komatsu:2008hk} we find that $(n_s{-}1) \sim O(10^{-2})$ at $q=0.002\, \mathrm{Mpc}^{-1}$, and hence  $g_{\mathrm{eff}}^2 \sim O(10^{-2})$ also, justifying our perturbative treatment of the QFT.

\begin{figure}[t]
\center
\includegraphics[width=0.45\textwidth]{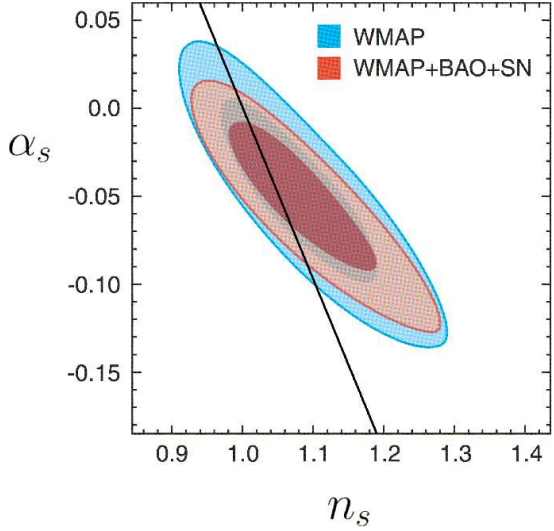} 
\hspace{2pc}
\begin{minipage}[b]{15pc}
\caption{\label{running_fig} The straight line is the leading order prediction of
holographic models with a single dimensionful coupling constant for the
correlation of the running $\alpha_s$ and the scalar tilt $n_s$.
The data show the $68\%$ and $95\%$ CL constraints (marginalizing over tensors)
at $q_0 = 0.002\, \mathrm{Mpc}^{-1}$, and are taken from Fig.~4 of \cite{Komatsu:2008hk}.
As new data appear the allowed region should shrink to a point,
which is predicted to lie close to the line.
\\
\\
}
\end{minipage}
\end{figure}

To determine whether the spectral tilts are red or blue requires evaluating the signs of $D_A$ and $D_B$, which will in general depend on the field content of the QFT.
It is nonetheless still possible to extract predictions which are independent of the field content:  for example, in these models, the scalar spectral index runs as
\[
 \alpha_s = \d n_s/\d\ln q =  -(n_s{-}1) + O(g_{\mathrm{eff}}^4).
\]
This prediction is qualitatively different from slow-roll inflation,
for which $\alpha_s/(n_s{-}1)$ is of first-order in slow-roll \cite{Kosowsky:1995aa}, yet is nonetheless consistent with the WMAP observational constraints on $n_s$ and $\alpha_s$ given in \cite{Komatsu:2008hk} for a wide range of values of $n_s$ and
$\alpha_s$, as illustrated in Fig.~\ref{running_fig}.

\subsection{Non-Gaussianities.}

Once $N$, $g_{\mathrm{YM}}^2$ and the field content are fixed, all
other cosmological observables (such as non-Gaussianities, {\it
 etc.})~follow uniquely from straightforward computations.
We will present details of
the correspondence between higher-order QFT correlation functions and
non-Gaussian cosmological observables elsewhere \cite{to_appear}.  Our
results indicate, however, that the non-Gaussianity parameter
$f_{NL}^{\mathrm{local}}$ \cite{Komatsu:2001rj} is independent of $N$
to leading order, consistent with current observational evidence
\cite{Komatsu:2008hk}.

\section{Conclusions.}  \label{sec:concl}

Let us summarize the main results. We have presented a holographic framework 
for early-universe cosmology describing the period of time 
corresponding to 
the inflationary epoch. 
In particular, we have shown how to compute cosmological observables 
by performing calculations with a three-dimensional dual QFT. 
This procedure was discussed explicitly for the case of the primordial power spectrum,
which is related to the 2-point functions of the dual QFT. Higher-point functions
are related to non-Gaussianities, as will be discussed elsewhere. When gravity is 
weakly coupled at early times, holography correctly reproduces
standard inflationary predictions for cosmological observables.
When gravity is instead strongly coupled at early times, one finds new models that have a
weakly coupled QFT description. 
We saw how models of this type exist that are compatible with current observations yet nevertheless have a distinct phenomenology from standard inflation.
The proposed holographic approach thus provides a qualitatively new method for generating
a nearly scale-invariant spectrum of primordial cosmological perturbations.

A special case to consider is de Sitter spacetime. A correspondence between de Sitter (dS) and CFT has been proposed in the past \cite{Strominger:2001pn},
and it is natural to wish to understand the relation of the present work to dS/CFT.
Under the domain-wall/cosmology correspondence, dS is mapped
to AdS, and our results for dS follow via suitable analytic continuation from 
the results for AdS. It is well known that quantum correlators, such as, for example,
the Feynman propagators of massive scalar fields in AdS and dS spacetime, do not map to 
each other under analytic continuation, see for example \cite{Das:2006wg}. This has been 
one of the obstacles in trying to establish a dS/CFT correspondence using analytic 
continuation. In our case, however, we do not map quantum de Sitter correlators to 
quantum AdS correlators. Instead, we map dS correlators directly to correlation functions
of the dual QFT, and (as we have seen) one can successfully establish such a map.
An approach to dS/CFT that has more in common with the present work is that of Maldacena \cite{Maldacena:2002vr},
although our analytic continuation is different: we analytically continue both the 
momenta and Newton's constant, whereas in \cite{Maldacena:2002vr} it was the dS radius
that was continued. Furthermore, compared to previous works \cite{Larsen:2002et,Larsen:2003pf,vanderSchaar:2003sz,Larsen:2004kf}
that focused mostly on the computation of the scalar power spectrum 
for asymptotically dS geometries, we gave a complete discussion of both the scalar and 
tensor power spectra and our work includes the case of asymptotically power-law inflation. Finally,
we proposed a precise definition of the dual QFT.

While uncovering an underlying holographic structure in inflationary cosmology
is conceptually important, this in itself would lead to no new hard results 
were one to remain in the regime where inflationary computations are well-justified. 
The reason is that in this regime, the holographic computations simply reproduce well-known results.
Rather, the power of the holographic approach is that it leads to 
new models in which the gravitational dynamics were strongly coupled at early times;
for these models the standard inflationary computations do not apply. 
This is precisely the distinguishing feature of the present work:
because we give an explicit definition of the dual QFT, we are able to obtain 
just such models with strongly coupled gravitational dynamics at early times.
These models have a weakly coupled dual QFT description, 
permitting the analysis of a scenario that would otherwise be quite intractable.

We have seen how the dual QFT may be defined in an operational sense by first performing all computations
with the ordinary QFT dual to a holographic RG flow, and then
continuing the number of colors $N$ and the momenta appropriately.
In the large $N$ limit the correlators are given as a series in $1/N^2$,
and so the analytic continuation simply amounts to inserting minus
signs. Furthermore, the continuation in momenta is such that the
effective coupling $g^2_{\mathrm{eff}} = g^2_{\mathrm{YM}} N/q$, where
$q$ is a momentum, remains real.  
It would be very interesting to understand such `pseudo'-QFTs from first principles. 
This would allow for a non-perturbative definition of the theory and would
presumably elucidate many puzzling features of quantum gravity
such as, for example, the entropy of de Sitter spacetime. Note 
also that the dual QFT provides a complete smooth 
description of the system, including that of the initial singularity that is
(generically) present in the background FRW solution. Our focus
here was to extract the late-time behavior of the system, but it would be 
very interesting to understand the implications for singularity 
resolution.

For the present, however, the main question is whether or not holographic models exist that
are compatible with current observations but have a distinct
phenomenology from standard inflation. We have shown that the answer
to this question is affirmative.  Initiating a holographic
phenomenological approach, we found that it is straightforward to
satisfy the current observational constraints using simple QFT models
containing only a few parameters.  The near scale invariance of the
cosmological power spectra follows immediately from simple dimensional
considerations.  A number of parameters in the QFT model may then be
estimated or constrained using the present observational data; once
the remaining parameters have been fixed, all cosmological observables
(including non-Gaussianities) then follow from direct
computation. Note that these are complete models: there are no UV
issues as these theories are super-renormalizable and furthermore the
dimensionful coupling constant acts as an infrared regulator.  Even
without knowing the values of all the parameters in the QFT model, it
is still perfectly possible to obtain concrete predictions, since not
all cosmological observables depend on the full set of QFT parameters.
Examples include the scale invariance of the power spectra at leading
order, as well as the running of the spectral index discussed above.
In general, we expect to obtain predictions that are qualitatively
different from those of standard inflationary scenarios based on
weakly coupled gravity.  These expectations are borne out by the form
of the running we found for the scalar spectral index.

Clearly, the proposed phenomenological approach to holographic
cosmology is worthy of further development.  Over the next few years,
forthcoming experiments (in particular the Planck satellite) promise
to dramatically improve the observational constraints on many
important cosmological parameters.  It may well be 
that future observations confirm 
the predictions of holographic models of
the type advocated here.  The success of such an endeavor might then
provide the first observational evidence for the holographic nature of
our universe.

\section*{Acknowledgments.}
We thank I.~Kanitscheider, B.~van Rees and M.~Taylor for useful discussions.
The authors acknowledge the support of NWO; KS through a VICI grant and PLM through a VENI grant.


\section*{References}

\begin{thebibliography}{99}

\bibitem{'tHooft:1993gx}
't Hooft G 1993 Dimensional reduction in quantum gravity  \pre gr-qc/9310026;
Susskind L 1995 The world as a hologram {\it J.~Math.~Phys.}~{\bf 36} 6377
  (\pre hep-th/9409089)

\bibitem{Maldacena:1997re}
Maldacena J M 1998 The large N limit of superconformal field theories and supergravity {\it Adv.~Theor.~Math.~Phys.}~{\bf 2} 231--52
(\pre hep-th/9711200)

\bibitem{Gubser:1998bc}
Gubser S S, Klebanov I R, and Polyakov A M 1998 Gauge theory correlators from noncritical string theory {\it Phys.~Lett.~}B {\bf 428} 105--14
(\pre hep-th/9802109)

\bibitem{Witten:1998qj}
Witten E 1998 Anti-de Sitter space and holography {\it Adv.~Theor.~Math.~Phys.}~{\bf 2} 253--91
(\pre hep-th/9802150)

\bibitem{McFadden:2009fq}
McFadden P L and Skenderis K 2009 Holography for cosmology \pre 0907.5542

\bibitem{Komatsu:2008hk}
Komatsu E {\it et al.}~2009 Five-year Wilkinson Microwave Anisotropy Probe (WMAP) observations: cosmological interpretation {\it Astrophys.~J.~Suppl.}~{\bf 180} 330 (\pre 0803.0547)

\bibitem{Turok:2002yq}
Turok N 2002 A critical review of inflation {\it Class.~Quant.~Grav.}~{\bf 19} 3449-67; 
Brandenberger R H 1999 Inflationary cosmology: progress and problems \pre hep-ph/9910410   

\bibitem{Maldacena:2002vr}
Maldacena J M 2003 Non-Gaussian features of primordial fluctuations in single field inflationary models {\it JHEP} {\bf 0205} 013 (\pre astro-ph/0210603)

\bibitem{Cvetic:1994ya}
Cvetic M and Soleng H H 1995 Naked singularities in dilatonic domain-wall spacetimes {\it Phys.~Rev.}~D {\bf 51} 5768 (\pre hep-th/9411170)

\bibitem{Skenderis:2006jq}
Skenderis K and Townsend P K 2006 Hidden supersymmetry of domain walls and cosmologies {\it Phys.~Rev.~Lett.}~{\bf 96} 191301 (\pre hep-th/0602260);
2007 Pseudo-supersymmetry and the domain-wall/cosmology correspondence {\it J.~Phys.~}A {\bf 40} 6733 (\pre hep-th/0610253)

\bibitem{Bergshoeff:2007cg}
Bergshoeff E A, Hartong J, Ploegh A, Rosseel J and van den Bleeken D
2007 Pseudo-supersymmetry and a tale of alternate realities JHEP {\bf 0707} 067 (\pre 0704.3559);
Skenderis K, Townsend P K and van Proeyen A 2007 Domain-wall/cosmology correspondence in AdS/dS supergravity JHEP {\bf 0708} 036 (\pre 0704.3918)

\bibitem{Skenderis:2008dh}
Skenderis K and van Rees B C 2008 Real-time gauge/gravity duality {\it Phys.~Rev.~Lett.}~{\bf 101} 081601 (\pre 0805.0150);
2009 Real-time gauge/gravity duality: prescription, renormalization and examples
  {\it JHEP} {\bf 0805} 085 (\pre 0812.2909)

\bibitem{Salopek:1990jq}
Salopek D S and Bond J R 1990 Nonlinear evolution of long wavelength metric fluctuations in inflationary models {\it Phys.~Rev.}~D {\bf 42} 3936

\bibitem{Bardeen:1983qw}
Bardeen J M, Steinhardt P J and Turner M S 1983 Spontaneous creation of almost scale-free density perturbations in an inflationary universe {\it Phys.~Rev.}~D {\bf 28} 679

\bibitem{Mukhanov:1990me}
Mukhanov V F, Feldman H A and Brandenberger R H 1992
Theory of cosmological perturbations {\it Phys.~Rept.}~{\bf 215} 203

\bibitem{DeWolfe:2000xi}
 DeWolfe O and Freedman D Z 2000
 Notes on fluctuations and correlation functions in holographic
  renormalization group flows \pre hep-th/0002226

\bibitem{Bianchi:2001de}
 Bianchi M, Freedman D Z and Skenderis K 2001
 How to go with an RG flow
{\it JHEP} {\bf 0108} 041
 (\pre hep-th/0105276)

\bibitem{Papadimitriou:2004rz}
Papadimitriou I and Skenderis K 2004 AdS/CFT correspondence and geometry
\pre hep-th/0404176;
2004 Correlation functions in holographic RG flows {\it JHEP} {\bf 0410} 075 (\pre hep-th/0407071)

\bibitem{to_appear}
McFadden P L and Skenderis K to appear


\bibitem{Skenderis:2002wp}
Skenderis K 2002 Lecture notes on holographic renormalization {\it Class.~Quant.~Grav.}~{\bf 19} 5849 (\pre hep-th/0209067)

\bibitem{Kanitscheider:2008kd}
Kanitscheider I, Skenderis K and Taylor M 2008 Precision holography for non-conformal branes {\it JHEP} {\bf 0809} 094 (\pre 0807.3324)


\bibitem{'tHooft:1973jz}
't Hooft G 1974 A planar diagram theory for strong interactions
{\it Nucl.~Phys.~}B {\bf 72} 461

\bibitem{Kanitscheider:2009as}
Kanitscheider I and Skenderis K 2009
Universal hydrodynamics of non-conformal branes
{\it JHEP} {\bf 0904} 062
(\pre 0901.1487)

\bibitem{de Haro:2000xn}
de Haro S, Solodukhin S N and Skenderis K 2001
Holographic reconstruction of spacetime and renormalization in the AdS/CFT correspondence
{\it Commun.~Math.~Phys.}~{\bf 217} 595 (\pre hep-th/0002230)

\bibitem{Boonstra:1998mp}
Boonstra H J, Skenderis K and Townsend P K 1999
The domain wall/QFT correspondence
{\it JHEP} {\bf 9801} 003
(\pre hep-th/9807137)

\bibitem{Papadimitriou:2005ii}
 Papadimitriou I and Skenderis K 2005
Thermodynamics of asymptotically locally AdS spacetimes
{\it JHEP} {\bf 0508} 004
(\pre hep-th/0505190)

\bibitem{xfig}
http://x-journals.com/2009/the-search-for-dark-energy-from-the-ground-up/

\bibitem{Jackiw:1980kv}
Jackiw R and Templeton S 1981 How super-renormalizable theories cure their infrared divergences {\it Phys.~Rev.}~D {\bf 23} 2291; 
Appelquist T and Pisarski R D 1981 High-temperature Yang-Mills theories and three-dimensional quantum chromodynamics {\it Phys.~Rev.}~D {\bf 23} 2305

\bibitem{Kosowsky:1995aa}
Kosowsky A and Turner M 1995 CBR anisotropy and the running of the scalar spectral index {\it Phys.~Rev.}~D {\bf 52} 1739 (\pre astro-ph/9504071)

\bibitem{Komatsu:2001rj}
Komatsu E and Spergel D N 2001 Acoustic signatures in the primary microwave background bispectrum {\it Phys.~Rev.}~D {\bf 63} 063002 (\pre astro-ph/0005036)

\bibitem{Strominger:2001pn}
Strominger A 2001 The dS/CFT correspondence {\it JHEP} {\bf 0110} 034 (\pre hep-th/0106113)

\bibitem{Das:2006wg}
Das A K and Dunne G V  2006
Large-order perturbation theory and de Sitter/anti-de Sitter effective actions
{\it Phys.~Rev.}~D {\bf 74} 044029
(\pre hep-th/0607168)


\bibitem{Larsen:2002et}
Larsen F, van der Schaar J P and Leigh R G 2002
de Sitter holography and the cosmic microwave background 
{\it JHEP} {\bf 0204} 047
(\pre hep-th/0202127)

\bibitem{Larsen:2003pf}
Larsen F and McNees R 2003 Inflation and de Sitter holography 
{\it JHEP} {\bf 0307} 051
(\pre hep-th/0307026)

\bibitem{vanderSchaar:2003sz}
van der Schaar J P 2004 Inflationary perturbations from deformed CFT
{\it JHEP} {\bf 0401} 070
(\pre hep-th/0307271)

\bibitem{Larsen:2004kf}
Larsen F and McNees R 2004
Holography, diffeomorphisms, and scaling violations in the CMB
{\it JHEP} {\bf 0407} 062
(\pre hep-th/0402050)




\end{thebibliography}
\end{document}